\documentclass[twocolumn,showpacs,preprintnumbers,showkeys,amsmath,amssymb]{revtex4}
\usepackage{graphicx}
\usepackage{dcolumn}
\usepackage{amssymb}
\usepackage{amsmath}
\usepackage{color}

\newcommand{\ud}{\mathrm{d}}

\begin{document}

\title{Harmonic moment dynamics in Laplacian growth}
\author{Alexander Leshchiner$^1$}
\author{Matthew Thrasher$^1$}%
\author{Mark B. Mineev-Weinstein$^2$}
\author{Harry L. Swinney$^1$}%
\email{swinney@chaos.ph.utexas.edu} \affiliation{ $^1\textrm{Center
for Nonlinear Dynamics and Physics Department, University of Texas at
Austin, Austin TX 78712 USA}$ \\ $^2\textrm{Applied Physics Division,
MS-P365, Los Alamos National Laboratory, Los Alamos, NM 87545 USA}$ }

\date{\today}

\begin{abstract}

Harmonic moments are integrals of integer powers of $z = x+iy$ over a
domain. Here the domain is an exterior of a bubble of air growing
in an oil layer between two horizontal closely spaced plates.  Harmonic
moments are a natural basis for such Laplacian growth phenomena because,
unlike other representations, these moments linearize the zero surface
tension problem (Richardson, 1972), so that all moments except the lowest
one are conserved in time. For non-zero surface tension,
we show that the the harmonic moments decay in time rather than
exhibiting the divergences of other representations. Our laboratory
observations confirm the theoretical predictions and demonstrate that
an interface dynamics description in terms of harmonic moments is physically
realizable and robust. In addition, by extending the theory to include surface
tension, we obtain from measurements of the time evolution of the harmonic
moments a value for the surface tension that is within 20\% of the accepted value.

\end{abstract}



\maketitle

\section{\label{sec:Intro}INTRODUCTION}

\subsection{\label{subsec:Laplacian Growth}Laplacian growth and viscous fingering}

Non-equilibrium processes give rise to a variety of patterns with remarkable geometrical
and dynamical properties \cite{CrossHohenberg,LangerGollub}.  Often these processes are
represented by the dynamics of an unstable interface between different phases, and the
interface patterns can exhibit universal features \cite{PelceBook}. Examples include crack
propagation \cite{LangerGollub}, fluid-fluid interface dynamics \cite{ST}, crystal
formation \cite{Langer}, and biological growth \cite{BenJacob}.


The simplest process leading to unstable universal patterns is {\em Laplacian growth},
where the velocity of an interface is proportional to the
gradient of function that is harmonic outside (or inside) the interface.
Here we examine the simplest example of Laplacian growth:
quasi-two-dimensional (2D) viscous fingering in a Hele-Shaw cell \cite{Hele-Shaw},
where a viscous
fluid is displaced by an inviscid fluid between two horizontal
closely spaced parallel plates.

\begin{figure}[htbp]
\includegraphics[width = 8cm] {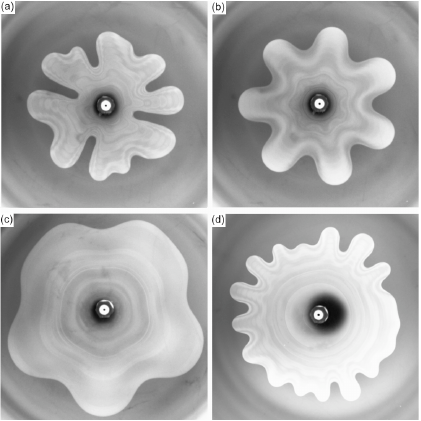}
\caption{Viscous fingering patterns of air in oil contained between
two plates separated by 125 $\mu$m (image size is 14 cm $\times$ 14
cm)  (a) Bubble grown at a constant
pumping rate of 0.52 mL/min. (b), (c), (d) Bubbles grown with varying
pumping rates, as described in Section IV.B. The time development of
bubble (b) is shown in Fig. 3.}
 \label{fig:Fig1}
\end{figure}

Figure \ref{fig:Fig1} shows four viscous fingering patterns grown in
the radial Hele-Shaw cell described in Section~\ref{sec:Experiment}.
Viscous silicone oil is removed from a buffer surrounding the
plates and air enters between the plates through a central hole in the
bottom plate.

As the air bubble expands, the oil/air interface is unstable. The
depth averaged velocity ${\bf v}(x,y)$ and the pressure $p(x,y)$ in the
oil are approximated by Darcy's law \cite{Lamb}
\begin{equation}
{\bf v} = - \frac{b^2}{12 \mu} \nabla p \, ,
\label{EqDarcy}
\end{equation}
where $b$ is the spacing between the plates and $\mu$ is the dynamic oil
viscosity.  The oil is incompressible, so ${\rm div}\,{\bf v} = 0$.
From (\ref{EqDarcy}) it then follows that in the oil

\begin{equation}
\nabla^2 p = 0 \, .
\end{equation}
Since the normal velocities of the interface $V$ and of the fluid at the
interface coincide \cite{Lamb},

\begin{equation}
V = - \frac{b^2}{12 \mu} \partial_n p \, , \label{Eq:Laplace}
\end{equation}
where $\partial_n$ is the normal component of the gradient. Because the
air is nearly inviscid, the pressure in the air is essentially
uniform, and its value can be taken as zero.  Thus the pressure jump
across the oil/air interface coincides with the oil pressure at the
interface and is given by \cite{ST}

\begin{equation}
p = - \sigma \kappa \, , \label{Eqpsk}
\end{equation}
where $\sigma$ is surface tension and $\kappa$ is the local curvature
of the interface in the horizontal plane. (An updated boundary
condition including a $\pi /4$ multiplicative factor and an additional
term correcting for wetting will be presented in Section
\ref{subsec:TheoryWetting}.)


The asymptotic pressure boundary condition in oil far from the
interface in the radial geometry is
\begin{equation}
p = -\frac{3\mu Q}{\pi b^2} \log(x^2 + y^2) \qquad ({\rm for} \,\,
x^2 + y^2 \rightarrow \infty), \label{Eqboundryatinf}
\end{equation}
where $Q$ is the pumping rate (the rate of area growth), which may
depend on time.  Equations (\ref{EqDarcy})-(\ref{Eqboundryatinf})
complete the description of 2D viscous fingering, which is a
prototype of Laplacian growth.

Laplacian growth, a classical free-boundary problem \cite{Oxf98}, is famous
in both physics and mathematics:  In physics, because the process
is highly unstable, dissipative, non-equilibrium, and nonlinear, and
because the growth produces universal patterns \cite{ST,Ristroph,Asympt}.
In mathematics, because the description (\ref{EqDarcy})-(\ref{Eqboundryatinf}),
as simple as it looks, reveals a powerful and profound structure \cite{MWZ,KMWZ,MPT}.

Laplacian growth occurs in many physical systems and is known by names such as
crystal growth, amorphous solidification \cite{Langer}, electrodeposition \cite{Gollub,Gollub2},
bacterial colony growth \cite{BenJacob}, diffusion-limited aggregation (DLA) \cite{DLA81},
motion of a charged surface in liquid Helium \cite{Zubarev}, and viscous fingering \cite{Praud}.
There are thousands of articles
(theoretical, experimental, and computational) devoted to Laplacian growth
if one includes work on closely connected problems such as the Stefan problem
(solidification) \cite{KKL,M-Kr}, DLA \cite{Halsey}, and the phase-field model \cite{LangerGollub}.
Nevertheless, many features remain unexplained, despite extensive effort and full
knowledge of the laws of physics describing the process.

\subsection{Harmonic Moments}

The present work concerns a powerful description of Laplacian growth
called {\it harmonic moments}.  Our work demonstrates that robust results
for the time evolution of harmonic moments can be obtained from viscous
fingering data.  Harmonic moments are defined as

\begin{equation}
M_k = \int_D z^{-k} \, \frac{dx dy}{\pi} = - \oint_{\Gamma} z^{-k}
\, \overline{z} \, \frac{\, d z}{2\pi i}, \label{Eq:Moments}
\end{equation}
where $k = 0,1,\ldots,\infty$, $z = x + iy$, and the domain of
integration $D$ is {\it exterior} to a pattern's boundary $\Gamma$
and bounded by a large circle on the outside.  These are exterior
moments, which are relevant for exterior Laplacian growth, when a
viscous fluid is outside the interface, as in this work. Interior
moments, which are relevant for interior Laplacian growth when a
viscous fluid is inside the interface, are defined similarly, but
with positive powers of $z$ under the integral, which is taken over
the interior of the pattern.

Harmonic moments are of fundamental importance for Laplacian growth
because in the absence of surface tension, there is an infinite number of conservation laws:
all moments $M_k$ (except for $k$ = 0) are {\it conserved in time}, as was discovered
by Richardson (1972) \cite{Richardson72} for the interior Laplacian growth problem. The
conservation laws have been extended to the exterior case \cite{M90} and are experimentally
confirmed in this work.  In the harmonic moments basis the whole evolution
of $D(t)$ is reduced to the time-dependent area of a growing bubble, which is $M_0(t)$.
This result selects the `harmonic basis' $M_k$ from all other bases as the
best basis for describing $D(t)$ in Laplacian growth.

Harmonic moments form a complete basis for representing any 2D
interface, regardless of its complicated shape, provided that $D$ is
analytic and singly connected \cite{NovStraVar,Var}.  Like Fourier
modes, each harmonic moment $M_k$ corresponds to a particular aspect
of the interface.  The moment $M_0$ is the area of the domain $D$ divided by
$\pi$.  Moments $M_k$ for $k \geq 1$ are proportional to the amplitude
of a monochromatic wave, which modulates the circle with an exactly $k$ waves
along the circumference,
$$z = R\,\exp(i \phi) [1+ a_k\,\, exp(- i k \phi)],$$
where $\phi$ is a `stream function' parameter along the interface, $a_k$ is an
amplitude, and $R$ is the bubble radius.  Specifically, $M_k = a_k\,R^{2-k}$.
For an approximately $n$-fold pattern, the
dominant moments are $M_n$, $M_{2n}$, $M_{3n}$, etc. [e.g. Fig.~\ref{fig:Fig1}(b)
is approximately
7-fold symmetric and Fig.~\ref{fig:Fig1}(c) is approximately 5-fold symmetric].
However, care should be taken
in comparing different moments since they have different units; the moment $M_k$ has
units cm$^{2-k}$.

Viscous fingering structures become extremely complex for high growth
rates, where tips repeatedly split and form new fingers and
fjords. Hence it is remarkable that theory predicts that a set of
purely geometrical quantities, $M_k$, will change slowly during the
growth process and will have a well-defined limit as the surface
tension approaches zero.  The unique
properties of harmonic moments have been already used
to establish connections between Laplacian growth and other fields of physics
and mathematics
\cite{MWZ,KMWZ,MPT}, but no previous studies have attempted to extract
harmonic moments from laboratory data. Our work demonstrates that robust
moments $M_k$ can indeed be obtained from experiments, and particularly
that the moments $M_k$ for $k \geq 1$ are conserved in the limit of zero surface
tension.

\subsection{Overview}

In the following section we present some additional properties of
harmonic moments.  Section III extends the theory for harmonic moments
in viscous fingering to the physical case, where the surface tension
is not zero. In this case, the moments $M_k$ with positive $k$ are no
longer conserved, but their time derivatives are proportional to surface tension and hence
vanish in the zero surface tension limit. Section IV presents our
experimental and data analysis methods. Section V presents results for
harmonic moments determined from experiments for a wide range of
conditions. Also we show that the results for the harmonic moments can
be used to deduce a physical parameter, such as surface tension. Section
VI discusses the significance of our results.

\section{HARMONIC MOMENTS: HISTORY AND APPLICATIONS}

\subsection{\label{subsec:IntroInversePotential}The Inverse Potential Problem}

The origin of harmonic moments dates back to Isaac Newton's study of
the Inverse Potential Problem, which was subsequently investigated by
Kelvin, Poincar\'{e} and many researchers in the twentieth
century.  The Inverse Potential Problem \cite{NovStraVar,Var}
asks how to recover
the shape of a domain occupied by a uniformly distributed mass, given
the gravitational potential created by this mass outside the domain.
Surprisingly, this classical mathematical problem was found to be at
the heart of the underlying mathematical structure of Laplacian growth
\cite{Richardson72,MWZ,KMWZ,MPT}.  In two dimensions the gravitational
potential created by uniformly distributed mass (with unit density)
occupying the domain $D$ is given by
\begin{displaymath}
\Phi(x,y) = \frac{1}{\pi}\int_D\,\log|z-z'|\,dx'dy' =
\end{displaymath}
\begin{equation}
 \frac{1}{\pi}\int_D\,\log|z'|\,dx'dy'-
{\cal R}e \sum_{k=1}^{\infty}M_{k} z^k/k,
\end{equation}
if $z$ and the origin lie outside $D$.  The $M_k$, defined in
(\ref{Eq:Moments}), are multipole moments of the mass distribution
\cite{Jackson}.  In Laplacian growth the $M_k$ are traditionally
called {\it harmonic moments}. In (\ref{Eq:Moments}) the moments
$M_k$ are defined for a radial geometry. For a rectangular geometry,
where matter occupies a horizontal semi-infinite strip $D$ with a
width $L$ bounded from the left by an arbitrary curve, and
periodically extended (repeated) both up and down infinitely, the
moments are defined as $M_k = \int_D \exp(- 2\pi k z/L) \, dx
dy/\pi$. This geometry is relevant for Laplacian growth in a
rectangular Hele-Shaw channel, which has been studied since the work
by Saffman and Taylor \cite{ST}.

\subsection{Connections and Applications}

In science and engineering it is often of interest to find the shape
of an object from the indirect measurement of the harmonic moments
$M_k$. The problem of domain reconstruction from its moments has
applications in many areas, including signal processing, probability
and statistics, tomography, and the inverse potential problem in
geophysics (magnetic and gravitational anomaly detection)
\cite{SanAntonio}.  While the recovery of shapes from moments is often
an ill-posed problem, it was recently recognized that the moments
problem allows the complete closed form solution for so-called
quadrature domains \cite{Gust-Put}, a branch of mathematics created in
1970s \cite{Aharonov-Shapiro,Sakai}.  Remarkably,
these solutions are based on a technique of numerical {\it linear}
algebra that yields numerically stable and fast algorithms and
exposes a deep connection between harmonic moments and a theory of
analytic approximation \cite{Milanfar1, Milanfar2}.

\section{\label{sec:Theory}THEORY OF HARMONIC MOMENTS WITH
NONZERO SURFACE TENSION}

\subsection{\label{subsec:TheoryDerivation}Derivation of $dM_k/dt$}

In Laplacian growth with nonzero surface tension the harmonic moments
are no longer conserved.  To relate the dynamics of the harmonic
moments to measurable quantities in viscous fingering, we now obtain
an expression for the time derivative of the harmonic moments in terms
of a line integral over the air/oil interface.  The time derivative of
$M_k$ follows from the definition (\ref{Eq:Moments}) and
(\ref{EqDarcy}):
\begin{displaymath}
\frac{d M_k}{d t} = \frac{b^2}{12 \mu}\oint_{\Gamma} z^{-k} \,
\partial_n p \, \frac{\, d l}{\pi} = \end{displaymath}
We want to exchange the derivative between $p$ and $z^{-k}$, so we add
and subtract $p\,\partial_nz^{-k}$ from the integrand,

\begin{equation} = \frac{b^2}{12 \mu}\oint_{\Gamma} (z^{-k} \,\partial_n p -
p\,\partial_nz^{-k} + p\,\partial_nz^{-k})\, \frac{\, d l}{\pi}.
\end{equation}
Here $l$ is the arc length along the interface $\Gamma$. The integral
of the first two terms is zero by Gauss's theorem, after using
(\ref{Eq:Laplace}) and the analyticity of $z^{-k}$ in $D$,

\begin{displaymath}
\oint_{\Gamma} (z^{-k}\,\partial_np - p\,\partial_nz^{-k}) d l =
\oint_{\Gamma} (z^{-k}\nabla p - p\nabla z^{-k})\cdot\vec{n} \, d l
\end{displaymath}

\begin{equation}
= \int_D \nabla\cdot(z^{-k}\nabla p - p\nabla z^{-k})\,dx\,dy = 0.
\end{equation}
Using (\ref{Eqpsk}) for the pressure, we have

\begin{equation}
\frac{d M_k}{d t} = - \frac{\sigma b^2}{12 \pi \mu}\oint_{\Gamma}
\kappa \,\,
\partial_n \, z^{-k} \,  dl \label{Eq:beforepressuresub}
\end{equation}
with the interfacial curvature given by $\kappa =d\theta/dl$, where
$\theta$ is the angle of a tangent line to the interface. Using the
identity $\partial_n \, z^{-k} = i\partial_l z^{-k}$ = $-i k z^{-(k+1)} dz/dl$,
where $\partial_l$ denotes a tangential derivative, we have

\begin{displaymath}
-\frac{\sigma b^2 i}{12 \pi \mu} \oint_{\Gamma} \frac{d\theta}{d
l}\,\,\partial_l \, z^{-k} \,  d l = \frac{k\sigma b^2i}{12 \pi
\mu } \oint_{\Gamma}\,\, z^{-(k+1)}\,\frac{d z}{d l}\, \, d \theta
\end{displaymath}

\begin{equation}
= \frac{k\sigma b^2}{12 \pi\mu}\oint_{\Gamma}\, \, z^{-(k+1)}\, d
\,e^{i\theta}.
\end{equation}
where the last expression was obtained using $dz/dl = \exp(i\theta)$.
Integrating by parts, we obtain
\begin{equation}
\frac{d M_k}{d t} = \sigma \, k(k+1)\,\frac{b^2}{12 \pi \mu}
\oint_{\Gamma}\,e^{i\theta}\, \frac{\ud z}{z^{k+2}}.
\end{equation}
Note that Richardson's result for the conservation of moments is
recovered in the limit that the surface tension $\sigma$ vanishes.
Surface tension is now a regular rather than a singular perturbation.

\subsection{\label{subsec:TheoryWetting}Wetting and scaling correction}

The pressure jump at the interface (\ref{Eqpsk}) must be modified because as the air bubble
advances between the glass plates, it leaves behind a wetting film on
each plate. The thickness of this film increases with the interface's
velocity \cite{ParkHomsy}. This film can be seen as interference
fringes on the images in Fig.~\ref{fig:Fig1}, because the interface is
slowing down as the bubble expands.  In addition, a factor $\pi/4$ is
needed in (\ref {Eqpsk}). The pressure jump including the wetting
correction and a factor $\pi/4$ [not in (\ref {Eqpsk})] was calculated
by Park and Homsy to be \cite{ParkHomsy},

\begin{equation}
p = - \sigma\left(\frac{\pi \kappa}{4}\right) + \frac{2
\sigma}{b}\left[1 + 3.8\left(\frac{\mu
V}{\sigma}\right)^{2/3}\right], \label{Eq:wettingcorrection}
\end{equation}
where $V$ is the local normal velocity of the interface.  The
thickness of the film predicted by Park and Homsy was experimentally
confirmed by Tabeling and Libchaber \cite{Tabeling} for $6 \times
10^{-4} < \frac{\mu V}{\sigma} < 3 \times 10^{-3}$.

The interface velocity $V$ is greatest at the finger tips and is much
smaller at the sides of the fingers. Conventionally the base of the
fjord is called a stagnation point, although in these experiments
there is small motion at the base of the fjords as a consequence of
relaxation due to surface tension; the base of the fjord becomes more
bulbous the longer surface tension has acted on it.

After substituting this new expression for the pressure jump into
(\ref{Eq:beforepressuresub}), we obtain

\begin{displaymath}
\frac{d M_k}{d t} = \sigma k \frac{b^2 \pi}{12\mu} \left[ \frac{k+1}{4}\, \oint_{\Gamma}\,e^{i\theta}\, \frac{dz}{z^{k+2}} \right.
\end{displaymath}

\begin{equation}
\left. + \frac{3.8\,i}{\pi b/2}\,\oint_{\Gamma}\,\left(\frac{\mu V}{\sigma}\right)^{2/3}\, \frac{dz}{z^{k+1}} \right].
\label{workingequation}
\end{equation}

\subsection{Testing the theory and determination of surface tension}

In Section V we compute the moments $M_k(t)$ for growing viscous
fingering patterns directly from the pattern geometry using the definition
(\ref{Eq:Moments}).  The time evolution of the moments can be compared to the result in the zero surface tension limit, where the moments are conserved.

We will also compare different moments by computing the normalized
amplitudes,
\begin{equation} a_k = |M_k|/|M_0|^{\frac{2}{2-k}}.
\label{Eq:Amplitude}
\end{equation}

Further, we can directly test the theory using (\ref{workingequation})
and measuring the time derivative of the moments and independently
determining:

\begin{enumerate}
\item the fluid surface tension $\sigma$,
\item the fluid viscosity $\mu$,
\item the thickness $b$ of the gap between the plates,
\item the interface velocity $V$ in the second integral in (\ref{workingequation}) (this can be obtained from analysis of images of the air bubble as a function of time,
described in Section \ref{Method:VSplines}), and
\item the two integrals in  (\ref{workingequation}).
\end{enumerate}

Alternatively, we can use the theory to determine a fluid property if
the other four quantities are independently known. This is the way in
which we determine the surface tension in Section V.  However, since
numerical differentiation of the data for $M_k(t)$ is difficult to do
accurately, we take one more step to obtain a working equation with
improved signal to noise: equation (\ref{workingequation}) is
integrated over a time interval to obtain

\begin{displaymath}
M_k(t_2)-M_k(t_1)=
\end{displaymath}

\begin{displaymath}
= \sigma k\frac{b^2}{48\mu}(k+1)\,\int_{t_1}^{t_2}
\oint_{\Gamma}\,e^{i\theta}\, \frac{d z}{z^{k+2}} d t
\end{displaymath}

\begin{equation}
+ \sigma k\frac{b^2}{12\mu\pi} \int_{t_1}^{t_2} \frac{3.8
\,i}{b/2}\,\oint_{\Gamma}\,\left(\frac{\mu
V}{\sigma}\right)^{2/3}\, \frac{d z}{z^{k+1}}d t.
\label{Eq:FinalInt}
\end{equation}

With only the definition of moments (\ref{Eq:Moments}), the dynamics cannot be related to known physical quantities. However, with (\ref{Eq:FinalInt}),  the dynamics can be quantitatively connected to known physical quantities.  Equation \ref{Eq:FinalInt} is a cubic equation of the form $a\sigma +
b\sigma^{1/3} + c = 0$, where $a$, $b$, and $c$ are complex numbers.
We solve this equation numerically in Section V.B to deduce the
surface tension and to test the predicted decay rates of the moments.

\section{\label{sec:Experiment}EXPERIMENT}

\subsection{\label{subsec:ExpApparatus}Apparatus}

An oil layer is contained in a Hele-Shaw cell consisting of two
horizontal, closely spaced glass plates with a hole through the center of
the bottom plate. When oil is pumped out of a buffer that surrounds
the oil layer, air (at nearly atmospheric pressure) enters the layer through
the hole in the bottom plate and forms a bubble in the center of
the oil layer.

The optically polished glass plates each have diameter 28.8 cm and
thickness 6.0 cm; each plate is flat to 0.2 $\mu$m, as described in
\cite{Ristroph,Praud}.  The gap between the plates was either 125
$\pm$ 5 $\mu$m or 384 $\pm$ 6 $\mu$m; most of the gap uncertainty
arises from using a micrometer to measure the thickness of the metal
shims that set the size of the gap. Interferometric measurements
using a sodium lamp showed that the gap thickness was uniform to 0.3
$\mu$m for the 125 $\mu$m gap and 1.6 $\mu$m for the 384 $\mu$m gap.

The oil was  Dow Corning 200 silicone oil at 24 $^\circ$C. We
measured the viscosity $\mu=49.9\pm 0.3$ mPa s with a Paar Physica
MCR300 rheometer. The surface tension $\sigma_{reference}=21.1\pm
0.1$ mN/m was measured by the Wilhelmy Plate method using a Kruss
K11 tensiometer. The density was measured to be $\rho = 0.9585 \pm
0.0005$ g/cm$^3$.

Bubble patterns were imaged from above with a CCD camera (1300 x 1030
pixels). The frame rate ranged from 1/6 to 2 frames/s, depending on
the growth rate of a bubble.

\subsection{\label{subsec:ExpGrowth}Growth of a bubble}

Experiments were initiated by obtaining a nearly circular air bubble,
grown by slowly withdrawing oil from the buffer surrounding the
gap between the plates.  After an initial nearly circular bubble had
been grown to a radius of at least 2 cm (to give good spatial
resolution in the images), multi-fingered bubbles like those in Fig. 1
were grown by using a precision computer-controlled syringe pump to
remove oil from the annular buffer at a specified rate. Growth of a
full-sized bubble (15-20 cm diameter) took from 30 to 1600 s
(typically 300 s).

\begin{figure}[htbp]
\includegraphics[width = 8cm] {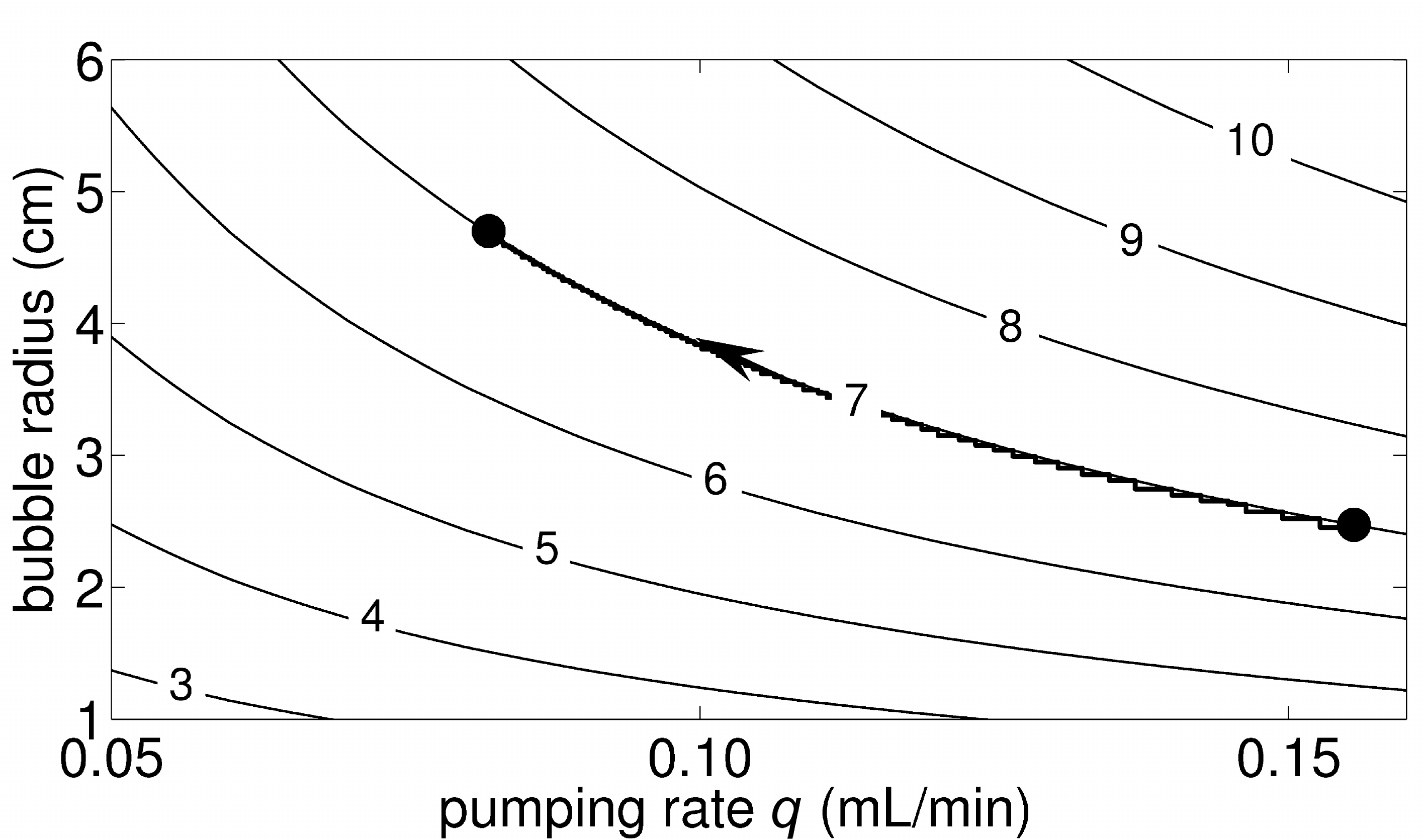}
 \caption{In order to grow an approximately $n$-fold symmetric bubble,
the pumping rate is adjusted to follow (\ref{Eq:Bataille}), as
illustrated here by curves for different $n$. The 7-fold bubble in
Fig. 1(b) was grown in this way, as indicated by the bold line. We
measure the area of a growing bubble in real time from the video
images, and compute from that area a radius for a circle with that
area. Then, given a target $n$-fold symmetry, the pumping rate is
adjusted (in small steps) to stay on the desired $n$-fold curve of
the graph. (These curves apply to a cell with gap thickness of 125
$\mu$m.)} \label{fig:FigQ}
\end{figure}

After obtaining a nearly circular bubble, a multi-fingered bubble can
be obtained by pumping oil out of the buffer. Usually fluid is
removed at a fixed rate, as in our laboratory's previous viscous
fingering experiments \cite{Ristroph,Praud}.  However, with a fixed
pumping rate, the $n$-fold mode that is the fastest growing changes
with time, and the resultant bubble has many azimuthal modes
with a substantial amplitude [e.g. Fig.~\ref{fig:Fig1}(d)].

Our procedure for obtaining the harmonic moments and deducing the
surface tension is applicable for bubbles grown with {\it any} pumping
rate, $q(t) =b Q(t)$.  However, results for the harmonic moments are
more robust if a bubble retains an approximate $n$-fold symmetry as it
grows.  To achieve this, the pumping rate must decrease in a way that
maintains an approximate $n$-fold symmetry. A linear
stability analysis by Bataille \cite{Bataille,Paterson} provides an
expression for the pumping rate needed to maintain the same $n$-fold
perturbation of a circle as the fastest growing \cite{Bataille,Paterson} mode,

\begin{equation}
Q_n = \frac{{\pi}\sigma b^{2}}{6\mu R}(3n^{2}-1)
\label{Eq:Bataille}
\end{equation}
where $R$ is the radius of the bubble. The bubble radius $R$ given by the Bataille formula is
plotted as a function of the volumetric pumping rate $q=bQ$ for different $n$-fold
modes in Fig.~\ref{fig:FigQ}. Conventional pumping with constant $Q$
would appear as a vertical trajectory in Fig.~\ref{fig:FigQ}.
For future experiments one should consider using Park-Homsy pressure jump,
$p=-\sigma \kappa \pi/4$, instead of using Bataille's assumption that $p=-\sigma \kappa$.


The particular multi-fingered pattern that develops depends not only
on pumping rate but also on initial perturbations and non-uniformities
of the glass plates. In the early stages of the pattern growth, when a
bubble begins to deviate from a circle, small unavoidable departures
from the circular shape have a greater influence on the development of
the interface than the accuracy of pumping or non-uniformities of the
glass plates.  In practice, we found the smallest $m$ achievable was
5; for smaller target $m$ the growth was too slow to overcome initial
perturbations on the bubble's surface before the bubble reached the
edge of the cell. We found that although the application of the
Bataille formula for small perturbations of a circular bubble is not
justifiable for bubbles that have grown large fingers, in practice the
Bataille's expression remains helpful in maintaining a symmetrical target
pattern.

The present work concerns bubbles that are growing throughout an
experiment. If the pumping were stopped, the surface tension would
begin to smooth the interfacial regions of high curvature, and an
interface could even reverse direction, absorbing the wetting film it
left behind.

In most viscous fingering experiments where oil is withdrawn at a
constant rate, the process has a natural separation of time scales:
$t_Q = A/Q$, determined by pumping rate, is usually much less than
the ``capillary'' time, $t_{\sigma} = 48\mu R^3/(\sigma b^2 n(n+1))$
(for not very high $n$), which corresponds to smoothing of a pattern
by surface tension. The difference between time scales ($t_{\sigma} >>
t_Q$) makes possible the rich interfacial patterns. However, in our
case there is only a single time scale because the pumping $Q$ is
adjusted continuously to maintain approximate $n$-fold
symmetry. Equating the scales $t_Q \thicksim t_{\sigma}$, one recovers
(within a multiplicative factor that depends on $n$) the Bataille
formula.  Note that for constant pumping the radius of a bubble is
given by $R \thicksim \sqrt{t}$, while in our case $R \thicksim
\sqrt[3]{t}$ (because $Q \thicksim 1/R$ and $R^2 \thicksim Qt$).

\subsection{\label{subsec:ExImAnalysis}Image analysis}

The oil/air interface in each image was first obtained by
subtracting the background image, thresholding, and using an edge
detection algorithm.  This located the interface to within a pixel.
The resolution of the interface was typically 50 pixels/cm,
so that one pixel's length was about 0.2 mm.

The position of each point on an interface was then obtained with
sub-pixel accuracy by interpolating the location along a line
perpendicular to the rough interface that was half of the intensity
difference between the inside of the bubble (more intense) and the
outside of the bubble (less intense). The algorithm typically found a
position of half-intensity to 0.1 pixel (about 20 $\mu$m, which is
small compared to the 125 $\mu$m or 384 $\mu$m plate separation).
This procedure yielded a sufficiently smooth interface so that
smoothing was not needed.

\subsection{\label{subsec:ExExtracting}Determing surface tension}
Equation (\ref{Eq:FinalInt}) is used to compute the surface tension.
The time $t_1$ is chosen to be when the amplitude $a_n$ [given by (\ref{Eq:Amplitude})] of the perturbation
from a circle exceeds 3 pixels.  The time $t_2$ is chosen so that for
slowly growing patterns about 10 images (60 s at 1 frame every 6 s) are
collected in $t_2 - t_1$; for rapidly growing patterns about 20 images
(10 s at 2 frame/s) are obtained.

The velocity $V$ was calculated by projecting the local normal to the
next later good interface. \label{Method:VSplines}  The use of splines allowed for the
intersection to occur between interface points, so that the velocity
could be computed more precisely.  Sums of points of the interface
were used as approximations of the contour integrals. The wetting
correction (the second integral) contains the capillary number, $Ca = \mu V
/ \sigma$, which in our experiments was in the range $ 9 \times
10^{-5} < \frac{\mu V}{\sigma} < 2 \times 10^{-2}.$  This range of $Ca$ is
slightly larger than the range that Tabeling and Lichaber observed the 2/3
power law relation of the film thickness [cf. (\ref{Eq:wettingcorrection})] \cite{Tabeling}.

\section{\label{sec:Results}RESULTS}
\subsection{Harmonic moments}

We have calculated the harmonic moments $M_k(t)$ by evaluating the
integrals in (\ref{Eq:FinalInt}) for 16 bubbles grown in a cell with
a 125 $\mu$m gap and for 10 bubbles grown in a Hele-Shaw cell with a
384 $\mu$m gap.  For each bubble the values of $M_k(t)$ were
computed for 50-500 data points spaced at time intervals 0.5-6 s.

We emphasize that our method for determining harmonic moments works
well for asymmetric as well as symmetric bubbles. We chose to attempt
to develop symmetric bubbles as they grew in order to track accurately
the time evolution of particular moments $M_k$. Hence for each bubble
the pumping rate was adjusted in real time according to
(\ref{Eq:Bataille}), as described in Section V.B (cf. Fig.~\ref{fig:FigQ}). Most
of the bubbles evolved toward a targeted $n$-fold symmetry, which
ranged from 5-fold (low pumping rate) to 14-fold (high pumping
rate). However, some bubbles had an initial shape that was too
irregular to evolve into an approximately $n$-fold bubble during the
course of the growth; Fig. 4 is an example of such a bubble.

\begin{figure}[htbp]
\includegraphics[width = 8cm] {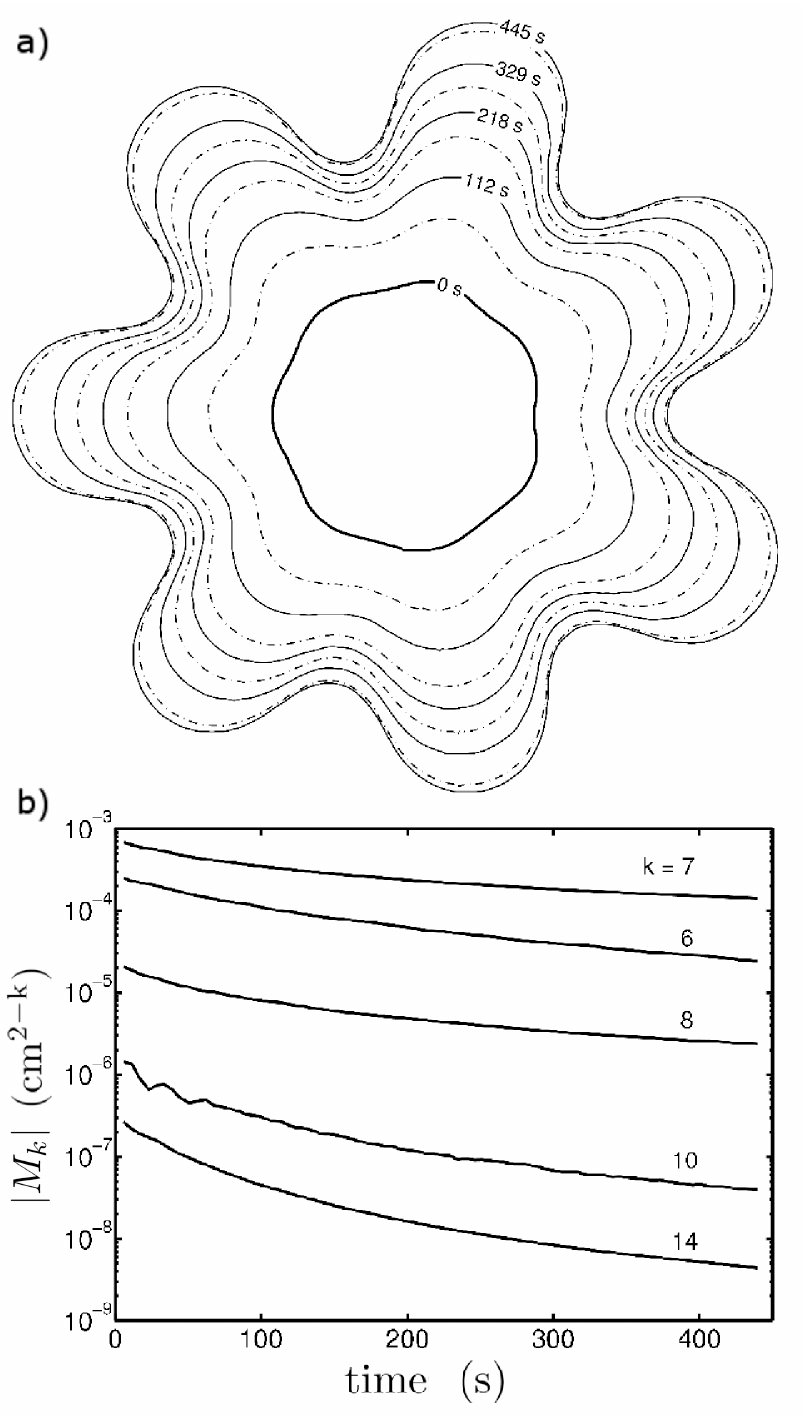}
 \caption{(a) An air bubble growing in oil in the Hele-Shaw cell (plate separation, 125
$\mu$m).  Adjacent interfaces are separated by 56 s, and the maximum
span of the bubble is 8 cm. The time-dependent pumping rate selects
the 7-fold symmetry (cf.  Fig. \ref{fig:FigQ}). (b) Amplitudes of
moments 6, 7, 8, 10 and 14, computed using (\ref{Eq:Moments}); note
that different moments have different units. All moments decay in
time, in contrast to the zero surface tension case where the moments
are constant.} \label{fig:Fig3}

\end{figure}

Our main result is that all observed moments $M_k(t)$ (except $M_0$)
decay in time, as Fig.~\ref{fig:Fig3} illustrates. The next subsection presents
results confirming that $dM_k/dt$ is proportional to the surface
tension $\sigma$ (neglecting wetting correction) [cf. (\ref{workingequation})], in accord with
Richardson's result that all moments are conserved in the zero surface
tension case.

\begin{figure}[htbp]
\includegraphics[width = 8cm] {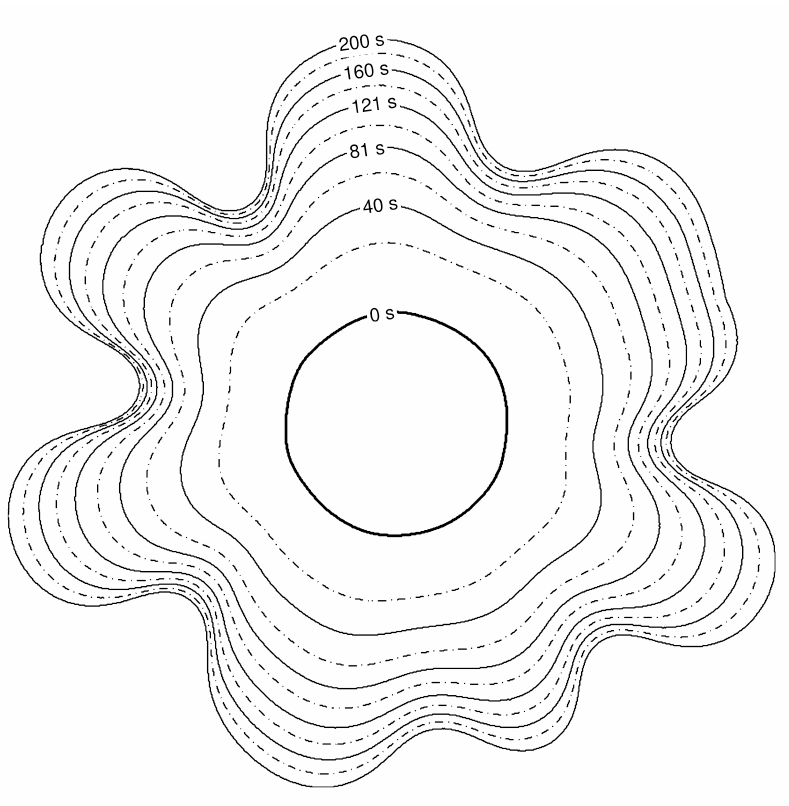}
\caption{An air bubble growing without a clean $n$-fold symmetry.
Adjacent interfaces are separated by 20 s. } \label{fig:Fig4}
\end{figure}

\subsection{Test of theory and determination of the surface
tension}

In the theoretical expression for $M_k(t_2)-M_k(t_1)$, equation
(\ref{Eq:FinalInt}), all quantities can either be determined from
our measurements or from independent measurements.  Hence the theory
can be directly tested with no adjustable parameters.

We chose to present the test of theory as the ratio of the
surface tension deduced from (\ref{Eq:FinalInt}), which we call
$\sigma_{measured}$, to the reference value of surface tension measured by
the Wilhemy Plate method, $\sigma_{reference}$. Thus surface tension is treated as an unknown whose value can be deduced
from (\ref{Eq:FinalInt}), where the integrals and velocity $V$
are determined from analyses of the bubble patterns, while the
viscosity $\mu$ and gap $b$ are measured independently.

\begin{figure}[htbp]
\includegraphics[width = 8cm] {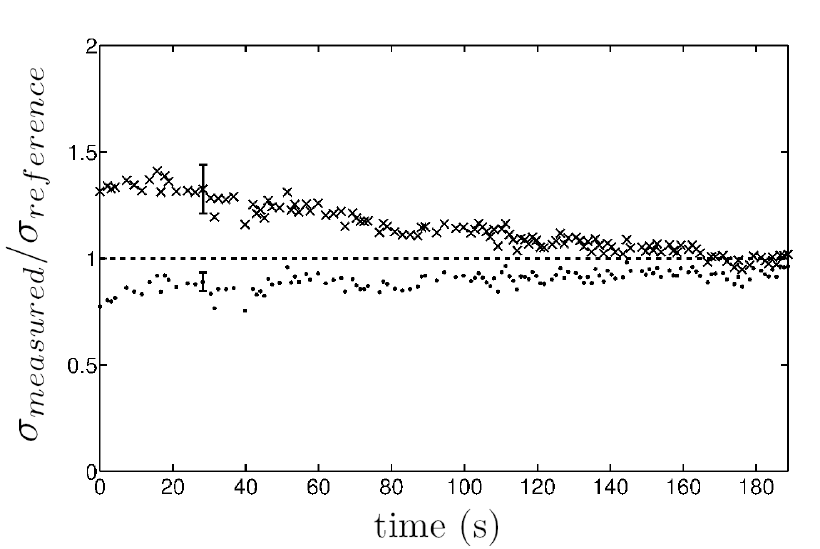}
\caption{ The ratio of the surface tension value deduced from theory
for the data in Fig. 4 to the value independently measured by
a traditional method (see text). The upper points ($\times$) are without a
wetting correction, and the lower points ($\bullet$) are with the wetting
correction in (\ref{Eq:FinalInt}) [the term with the factor 3.8].
The error bars show the standard deviation for each group of points.
The value of $t_2 - t_1$ in (\ref{Eq:FinalInt}) was 10 s. The horizontal axis corresponds to $t_1$.}
\label{fig:Fig5}
\end{figure}

An example of the results deduced for the surface tension ratio is
shown in Fig.~\ref{fig:Fig5}, which was computed for the non-symmetric bubble
pattern in Fig.~\ref{fig:Fig4}.  For this bubble, the surface tension deduced
without a wetting correction is 30\% larger than the reference value at short times, where
the front velocity is large (see Fig.~\ref{fig:Fig4}), while at long times the
front velocity has become smaller and the difference between the
ratios with and without the wetting correction is only a few percent.
Further, both ratios are within a few percent of unity, so the
value of surface tension deduced from theory is equal to the
reference value within the experimental uncertainty.

\begin{figure}[htbp]
\includegraphics[width = 8cm] {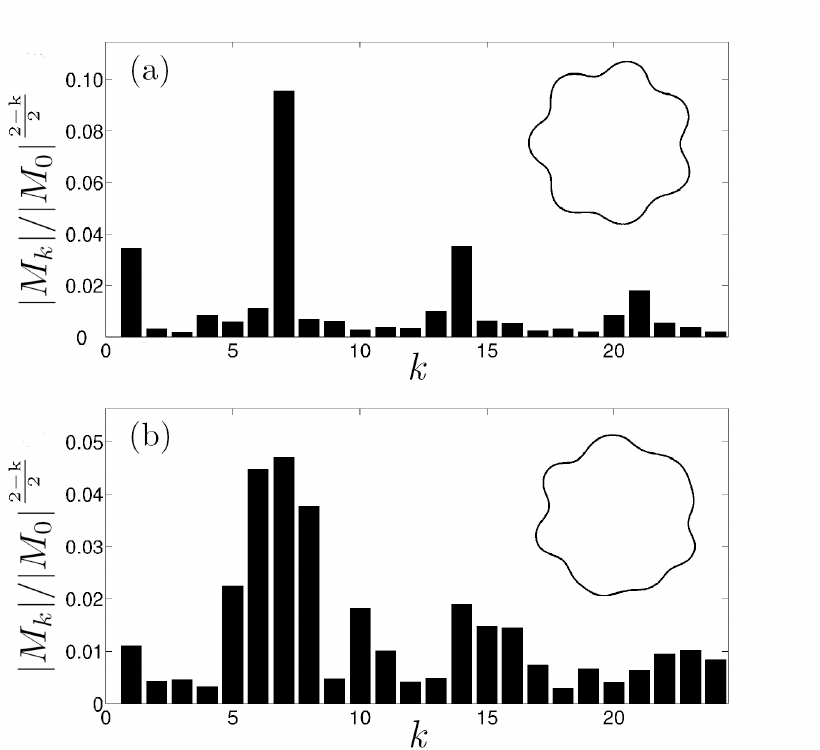}
 \caption{Spectra of the nondimensional moment amplitudes $|M_k|/|M_0|^{\frac{2-k}{2}}$
[equation (\ref{Eq:Amplitude})] for two bubbles: (a) A bubble with
approximately 7-fold symmetry [Fig. \ref{fig:Fig3}(a)], where the
main moments are 7, 14, and 21. The moments were calculated at the
112th second of the experiment. (b) A bubble without $n$-fold
symmetry (Fig. \ref{fig:Fig4}), where there is a broad spectrum with
no dominant moment. The moments were calculated at the 64th second
of the experiment.}

\label{fig:Fig6}
\end{figure}

A measure of the symmetry of a bubble is given by the spectrum of the
harmonic moments, i.e. a plot of the normalized moment amplitude, as
shown in Fig.~\ref{fig:Fig6} for the symmetric bubble in Fig.~\ref{fig:Fig3} and for the
non-symmetric bubble in Fig.~\ref{fig:Fig4}. The approximately 7-fold bubble in
Fig.~\ref{fig:Fig3} has a spectrum with significant components at 7, 14, and 21
[Fig.~6(a)], while the the non-symmetric bubble in Fig.~\ref{fig:Fig4} has a
spectrum with a larger number of significant components, including
those at 5, 6, 7, 8 [Fig.~\ref{fig:Fig6}(b)].

The mean value for the surface tension $\sigma_{measured}$ was
deduced from (\ref{Eq:FinalInt}), including the wetting correction,
for each of the 26 bubbles studied, where each bubble was evaluated
at 50-500 times during the course of its growth; this involved the
numerical evaluation of the integrals in (\ref{Eq:FinalInt}) for
more than 6000 data sets. The result is $\sigma_{measured}$ =
18$\pm$4 mN/m, where the uncertainty includes both the statistical
uncertainty and the estimated systematic uncertainty (see next
paragraph). The result for $\sigma_{measured}$ agrees within the
experimental uncertainty with $\sigma_{reference} = 21.1\pm0.1$
mN/m, determined by the Wilhelmy Plate method.  Thus the theory is
quantitatively confirmed within the experimental uncertainty.  [If
the wetting correction in (\ref{Eq:FinalInt}) is neglected, the
result for the surface tension is $23\pm6$ mN/m.]

\begin{figure}[htbp]
\includegraphics[width = 8cm] {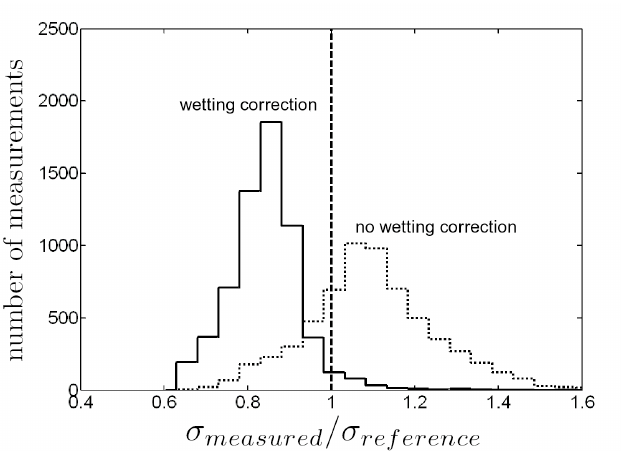}
 \caption{The distribution of the ratios of surface tension deduced from theory,
$\sigma_{measured}$, to $\sigma_{reference}$, determined by the Wilhelmy Plate method.
The solid line represents computations that include the wetting term, while the
dotted line neglects the wetting correction. The bin size is 0.05. A
  few outlying points are outside of the abscissa's range.}

\label{fig:Fig7}
\end{figure}

The uncertainty in our result for $\sigma_{measured}$ unfortunately
arises in part from a possible systematic error of 9\%. The
uncertainty in gap thickness (about 4\%) contributes significantly
to the overall uncertainty. Also, an intermittent problem in camera
synchronization could have introduced timing errors in some of the
data.  Other possible sources of error include the discretization of
the integrals and the approximations introduced by the theoretical
analysis. The range of applicability of the 2/3 scaling for the film
thickness is smaller than the range of $Ca$ produced in our
experiments. Another intriguing possible source of error is the 3.8
factor in the wetting correction \cite{ParkHomsy}, which has its
origin in a numerical integration done in the 1974 PhD Dissertation
by Ruschak \cite{Ruschak}. If the numerical factor of 3.8 were
changed to 1.5, then the mean of the distribution in
Fig.~\ref{fig:Fig7} would correspond to unity. Note that the wetting
correction becomes small at long times because the growth velocity
$V$ becomes small if the bubbles are grown according to
(\ref{Eq:Bataille}).

The statistics of the results are presented in Fig.~\ref{fig:Fig7}
as histograms of the ratio $\sigma_{measured}/\sigma_{reference}$,
calculated both with and without the wetting correction.  The
distribution including the wetting correction is narrower because
the uncorrected data depend on the pattern growth velocity.  The
mean value of the ratio with wetting correction is 0.84$\pm$0.09
(standard deviation, statistical uncertainty only); the mean value
without wetting correction is 1.10$\pm$0.1(standard deviation).

Future experiments can straightforwardly reduce the systematic uncertainty by
an order of magnitude to a level approaching 1\%. Such experiments should focus on low order moments at
intermediate times, to minimize the correction due to wetting and prevent backward motion of the interface.

Bubbles in our experiments were all studied for pumping rates $Q > 0$.
Alternatively, one could stop the growth (set $Q$=0) and observe
the relaxation of a bubble. We conducted a few experiments in this way
and found that the extracted surface tension values were typically 15-25\%
lower than those for growing bubbles. However, for relaxing bubbles
the correction for the oil wetting film, which is re-absorbed as the
interface retreats, is not known.

\section{\label{sec:Discussion}DISCUSSION}

\vspace{.1in}

\subsection{Harmonic moments}

We have presented the first demonstration that robust harmonic moments
$M_k(t)$ can be deduced from laboratory experiments. The $M_k(t)$
evolve in time because of nonzero surface tension [Fig. 3(b)];
otherwise all moments except the area $M_0$ would be conserved.  The harmonic moments are purely geometric quantities [cf. (\ref{Eq:Moments})] and do not involve approximations, fluid properties, or experimental parameters, such as the wetting correction, viscosity, or the gap thickness.

Further, we have obtained an equation for the dynamics of the harmonic
moments, equation (\ref{Eq:FinalInt}), which involves quantities that
can all be directly determined by experiment, thus providing a
quantitative test of the theory of harmonic moments.  The result from measurements on 26
bubbles for two different Hele-Shaw cell gap thicknesses is that
theory and experiment agree within the experimental uncertainty (about
20\%). This agreement was obtained both on nonsymmetric bubbles and on
bubbles with approximate $n$-fold symmetries varying from 5 to 14.
This agreement is robust and, within our experiments, does not depend on how far a bubble has
deviated from a circle. Further, we have indicated how it should be
straightforward to reduce the experimental uncertainty by an order of
magnitude, to test the theory at about a 1\% level.

For convenience we have used an expression obtained by Bataille
(\ref{Eq:Bataille}) to adjust the pumping rate $Q(t)$ to grow bubbles
that are approximately $n$-fold symmetric.  We find that if an
$n$-fold bubble is grown initially from a slightly perturbed circle,
the $n$-fold symmetry is retained even for large amplitude fingers,
far beyond range of applicability of (\ref{Eq:Bataille}), which was
obtained from a linear stability analysis.

\subsection{Implications}

Harmonic moments $M_k$ form a complete basis for any complicated time
domain $D$, provided that $D$ is analytic and singly connected. These
moments are exceptionally useful for Laplacian growth phenomena such
as viscous fingering because all other known representations involve
coefficients that change quickly because of instability, thus creating
analytic and computational difficulties. In contrast, harmonic moments
are free of these problems, no matter how much the domain $D$ deviates
from a circle. Loosely speaking, for harmonic moments
the whole instability is simply concentrated in the lowest one, $M_0$
(the area of $D$).

From a fundamental perspective, the moments $M_k$ for Laplacian
growth are exactly the basis that linearizes and ``solves'' the zero
surface tension problem, generating a multitude of exact solutions
(summarized in \cite{MPT}; see also references therein), which are
impossible to
obtain in any other basis. In this sense the harmonic moments constitute the most natural
basis for Laplacian growth. Our experiments confirm that for nonzero
surface tension, the derivative of the moments $dM_k/dt$ is simply proportional to
$\sigma$ (neglecting the wetting correction), so that the zero surface tension limit is
smoothly approached.  This indicates that the fundamental theoretical results mentioned
above, which have been obtained for Laplacian growth theory for the
case of zero surface tension, are relevant to real physical systems.

In conclusion, the implications of the demonstration of the harmonic moments description
of viscous fingering extends into various branches of mathematical physics (cf. Section II)
because of the deep connection between Laplacian growth and other problems.

\section{\label{sec:Acknowledgements}ACKNOWLEDGMENTS}
We thank Olivier Praud for developing the method for maintaining an
approximate $n$-fold symmetry of a bubble (Section IV.B), and we thank
Dmitry Leshchiner for helpful discussions. Acknowledgment is made to
the Donors of the American Chemical Society Petroleum Research Fund
for support of this research.  This work was also supported in part by
the LDRD program 20070083ER at Los Alamos National Laboratory.


\end{document}